\documentclass[9pt, a4paper, 
superscriptaddress
]{article}  

\usepackage[utf8]{inputenc} 

\usepackage{geometry}
\usepackage[english]{babel}
\usepackage{csquotes}
\usepackage{multicol}
\usepackage[
backend=biber,
style=ieee,
sorting=none,
citestyle=ieee
]{biblatex}
\usepackage{authblk}
\usepackage[hyperindex,breaklinks]{hyperref}

\hypersetup{
    colorlinks=true,
    linkcolor=blue,
    filecolor=blue,      
    urlcolor=blue,
    citecolor=blue,
    bookmarksopen=true,
    pdfpagemode=FullScreen,
}
\usepackage[anythingbreaks]{breakurl}

\usepackage{titlesec}
\titleformat{\subsection}
  {\normalfont\small\bfseries}
  {\thesubsection}{5pt}{}

\titleformat{\section}
  {\large\bfseries\scshape}{\thesection}{5pt}{}

\usepackage[capitalize]{cleveref}
\usepackage{enumitem}

\usepackage{epigraph}
\usepackage{etoolbox}

\makeatletter
\patchcmd{\epigraph}{\@epitext{#1}}{\itshape\@epitext{#1}}{}{}
\makeatother

\title{\large \textbf{Systems of Global Governance in the Era of Human-Machine Convergence}}

\author[1]{Eugenio Maria Battaglia}
\affil[1]{\small{\emph{Collective Intelligence Cluster,\space International Hackteria Society,\space Zürich, Switzerland\\}}}

\author[2]{Jie Mei}
\affil[2]{\small{\emph{Department of Experimental Neurology,\space Charité – Universitätsmedizin Berlin,\space Berlin, Germany}}}

\author[3,4]{Guillaume Dumas}
\affil[3]{\small{\emph{Human Genetics and Cognitive Functions Unit,\space Institut Pasteur,\space Paris, France}}}
\affil[4]{\textbf{Corresponding author:} \space \emph {Human Genetics and Cognitive Functions Laboratory\\Institut Pasteur, 25 rue du Docteur-Roux, 75015 Paris, France – \space}
\href{mailto:guillaume.dumas@pasteur.fr}{guillaume.dumas@pasteur.fr}}
\begin{document}
\maketitle
\begin{abstract}
\small Technology is increasingly shaping our social structures and is becoming a driving force in altering human biology. Besides, human activities already proved to have a significant impact on the Earth system which in turn generates complex feedback loops between social and ecological systems. Furthermore, since our species evolved relatively fast from small groups of hunter-gatherers to large and technology-intensive urban agglomerations, it is not a surprise that the major institutions of human society are no longer fit to cope with the present complexity. In this note we draw foundational parallelisms between neurophysiological systems and ICT-enabled social systems, discussing how frameworks rooted in biology and physics could provide heuristic value in the design of evolutionary systems relevant to politics and economics. In this regard we highlight how the governance of emerging technology (i.e. \emph{nanotechnology, biotechnology,
information technology, and cognitive science}), and the one of climate change both presently confront us with a number of connected challenges. In particular: historically high level of inequality; the co-existence of growing multipolar cultural systems in an unprecedentedly connected world; the unlikely reaching of the institutional agreements required to deviate abnormal trajectories of development. We argue that wise general solutions to such interrelated issues should embed the deep understanding of how to elicit mutual incentives in the socio-economic subsystems of Earth system in order to jointly concur to a global utility function (e.g. avoiding the reach of \emph{planetary boundaries} and widespread social unrest). We leave some open questions on how techno-social systems can effectively learn and adapt with respect to our understanding of geopolitical complexity.
\begin{description}
\item[Keywords]
\emph{NBIC, complex adaptive systems, sociophysics, digital platforms, governance.}
\end{description}
\end{abstract}
\medskip  
\begin{multicols}{2}
\setlength{\parindent}{5pt}
\setlength{\parskip}{1pt}
\section{Introduction}
This note is organized as follows. We begin with a careful recapitulation of how technology has shaped human history taking into consideration the interactions and analogies between biotic systems (in particular humans) and machines. Here we discuss the emerging and converging technologies challenging the distinction between humans and machines: a relation known as the \emph{nanotechnology, biotechnology, information technology, and cognitive science} convergence (NBIC).  \par Then we reflect on how social systems that diverge into cultural and technological backgrounds could coexist in an increasingly connected world. In particular, we explore how–and if–we could set the design of techno-social systems that could keep life-supporting systems at a regime that is compatible with life, thus ultimately avoiding the extinction of our species. We argue that there’s a rising need for a better understanding of the individual, social, ecological, and evolutionary consequences (i.e. opportunities and threats) of massive technology adoption, as well as the urgent need to broaden the current debate around NBIC technologies to a wider audience. As an example of such an interdisciplinary approach, we draw foundational parallelisms between neurophysiological systems and ICT-enabled social systems.We conclude that such kind of framework can provide heuristic value in the design of evolutive governance systems relevant to politics and economics. In particular, we argue that bio-inspired organization designs can answer some of the many ethical and governance issues raised by a complex and heterogeneous human society living on a planet heading to a global ecological state shift. \par We leave some open questions on how techno-social systems can effectively \emph{learn} and \emph{adapt} with respect to our understanding of geopolitical complexity. This note is meant to represent neither a complete survey of the complexity underlying human systems, nor of their technological landscape, nor of the governability of socioeconomic systems. Our aim is to convey a series of urgent and interconnected issues to spark meaningful conversations and collaborations around these topics. Towards the conclusion of the manuscript we specify some lines along which–in our opinion–this research could further develops.

\subsection{The economy of atoms and societies}
Self-organization is a property that has been observed in a variety of natural and artificial systems across scales \cite{Feistel2011,Karsenti2008,Nicolis1977,Wennekamp2013,Wollman2013}. Atoms find an energetic stability in chemical bonds \cite{Pauling1960}; biological replicators find a convenient diversity in reproduction \cite{Pachepsky2001,Szathmary1997}; biological cells are organized in intracellular compartments and organize into tissues for the specialization of biological activities \cite{Cremer2001,Luisi2016}; biological organs are organized into an individual being \cite{Bryant2008, Sober1991}; and, finally, biological individuals self-organize into social groups that possess emergent behaviors from cooperative and competitive interactions \cite{Adami2017,Ellwardt2016,Holling2001,Parrish1999,Pereda2016}. \par In a widely known essay dating back to 1944, Schrödinger \cite{Schrodinger1948} already noted that high level structures are optimal to withstand random dissipation in molecular systems. Indeed, starting from the intrinsic nature of hybridized carbon to organic and biochemical systems up, many structural configurations emerged as intermolecular and supramolecular interactions to delocalize the issue of stability to higher-order levels. This trend---as already observed by Schrödinger--–seems to be maintained as well in biological systems wherein further high-order structures emerge in order to increase the time needed for the overall system to relax to thermal equilibrium.\par The activities of human society, for instance, depend on our ability to constantly transform available sources of energy into work. In fact, most systems found on our planet, natural or engineered, are not at thermodynamic equilibrium since they are exchanging matter and energy with their surroundings, and change---or can be triggered to change---over time.\par Albeit, biological and socioeconomic systems are both far from equilibrium \cite{BrianArthur2013, Reiss1994}, and thus can be studied from this lens, mainstream neoclassical economic theories are generally not grounded into physics. As such, most economic theories are---for instance---not consistent with empirical evidences of scarce resources out of which living systems in the biosphere extract low entropy to compensate for their continuous dissipation. Nonetheless, the understanding of socioeconomic systems from the perspective of energy, material and information flows is rooted in anthropology \cite{Cronon1983,Debeir1991,Perlin2005,Tainter1990} as well as in economic analysis \cite{Chakrabarti2006,Chen2016,Ruth1993}.\par

\subsection{Coevolutionary dyanmics in Earth system}A series of efforts have already been made to schematize from different perspectives the historical transitions of the \emph{geochemical and living systems} on Earth \cite{Calcott2011,DeDuve2005,Knoll2000,Szathmary2015}. These studies have provided us with many insightful considerations, among which are the following: some recurrent patterns of evolutionary dynamics and what the relationship is between evolutionary trajectories and ecological complexity; what is required and expected by each transition in terms of free energy and material input/output, linking to the sustainability of specific regimes; at which levels natural selection operates (from genes to individuals and so on). The energetic breadth of our activities on this planet influences many dimensions of Earth’s ecology dynamics, which in turn mark the potential evolutionary paths of our species. As Donges and colleagues argue: \emph{“Understanding and modeling the Anthropocene, the tightly intertwined social-environmental planetary system that humanity now inhabits, requires addressing human agency, system-level effects of networks and complex coevolutionary dynamics.”}\cite{Donges2017}\par Indeed, since the speciation of \emph{Homo sapiens} that occurred between 200,000 and 400,000 years ago in Africa \cite{Brauer2008,Hublin2017}, the behavior of our species has dramatically changed in relation to environment, technology, and social interactions \cite{Kuhn2004}. The main behavioral changes occurred in the last 10,000 years due to Earth’s remarkably stable environment \cite{Kuper2006,Steffen2015,Ullah2015}. This period marks the beginning of an era that has allowed humans to spread into complex and diverse societies, exploring rich ecological niches, and develop technologies that have further significantly impacted the evolution of our behaviors and social structures.\par
Technology has indeed helped many civilizations cope with challenging environments and eventually produce a diverse set of cultural and biological traits \cite{Taylor2010}. The Stone Age became the starting point for hominins to manufacture stone tools, such as the hand axe \cite{Henshilwood2002}. From then on, the usage of tools was gradually woven into the evolutionary history of mankind. However, most of our technological development, demographic expansion, and consequent socio-cultural changes happened in the last 200 years, most likely due to low-priced potential energy in the form of fossil hydrocarbons. Such an unprecedented growth rate—which is unlikely to significantly decrease \cite{Brown2011,Burger2012,Burger2017,Judson2017,Obama2017}—has led to technological advancements and societal transformations that progressively amplified the individual and collective capabilities of our species. As clearly argued by Judson \cite{Judson2017}, the interplay between evolving life forms and the transformation of their planetary home is crucial to determine the development of the life-planet systems, a process that is highly dependent on the paths that evolving life can potentially take.\par

\subsection{The rising tide} The norms, agreements, cultural codes, and ways of organizing currently in place in our society's major institutions seem not fully mirroring the pace at which human behaviors emerge and evolve from small groups of hunter-gatherers to vast technology-intensive urban agglomerations. This inherent lag of \emph{formal systems} (i.e. norms, agreements, cultural codes, and ways of organizing) to properly manage, govern and cope with what we can define as \emph{real systems} (e.g. the set of behaviors and technologies in hunter-gatherers or in urban agglomerations), may be due to several reasons. First, the intuitive fact that the confidence of institutions in formalizing consistent set of rules around any matter would only increase with time (i.e. after having collected a considerable amount of observations). In fact, whether a policy, an agreement, or a norm is beneficial and well-pondered can only asserted a posteriori, as time pass.\cite{rovelli2017} Additionally, understanding the lag of formal systems in representing real systems may also be matter of observing the discrepancy between the cumulative nature of how formal systems are produced and the emergent, unforeseeable nature of the behaviors of real system.\par If in time these considerations will result adequate, it will be accordingly pertinent to say that there is an unavoidable \emph{governance deficit} for which institutions and human organizations would always lag in their ability to cope with emergent complexity. Ultimately, a governance deficit in a domain would be influenced by the domain's historical dependencies and contextual specificities. 

\subsection{An unprecedented technological landscape}One of the key objective of this manuscript is to direct attention towards the current capabilities of our society’s governance structures, questioning whether political institutions, mulilateral systems integrators (e.g. UN System agencies) or other traditional forms of ethical deliberation are fit to cope with the complex and delicate balance of our life-planet system, and with the emergence of an unprecedented technological landscape. We give particular focus to science and technology that are now increasingly able to act from the nano scale of our bodies to the global scope of our societies. Their impact brings forth an urgent call for ethical reflections \cite{Boenink2010,Cobb2004,Nordmann2009,Rogers-Hayden2007}. For instance, technologies like genetic engineering \cite{Cong2013} and neural implants \cite{Luan2017} are challenging the boundaries of biological individuality. Concurrently, other technology-enabled abilities like internet-mediated collective organization \cite{AnderssonSchwarz2017} and data-driven policy simulation are challenging the notion of human agency \cite{Bollier2010,Edmonds2013,Lafuerza2016,Wang2016}. By amplifying and diversifying humans’ decision-making processes and execution capabilities, we argue that these novel socio-technical systems---along with the inherent cultural load they carry \cite{Richerson2013}---are providing the fundamental requirements for an unprecedented reorganization of human social structures at different scales.\par Therefore, another key objective of this manuscript is to highlight the unclear impact of these emergent socio-technical systems and their interplay with a series of core geopolitical drivers. We concur with other authors that some further logical developments of our history might see the rapid dissolution of current political and socio-economical structures in favor of distributed systems of delegation and techno-mediated collective organizations \cite{AnderssonSchwarz2017,Davidson2016,Tapscott2017,Wright2015}. Indeed, today’s technological landscape is inviting widespread deliberations on the ways in which humans can and should use science and technology to manage their activities on Earth to avoid reaching planetary boundaries and provoking unpredictable consequences at the socio-political level (e.g. diffuse social unrest) \cite{Hsiang2017,Jaumotte2013,Stern2007}.\par As already mentioned a theme of particular relevance for this research revolves around the progressively growing interaction between human and machines through the convergence of nanotechnology, biotechnology, information technology, and cognitive sciences---a relation known as NBIC convergence. This notion has been first popularized in 2001 during a seminal workshop organized by the National Science Foundation (NSF), a U.S. government agency \cite{Roco2003}. The NBIC convergence relates to the embodied---even implanted---interactions between humans and artificial systems through \emph{“additional sensors and augmented reality”} \cite{Viaud-Delmon2012} that can significantly enhance people’s health, cognitive or physical capacities, as well as the competitive, operational, and employment landscape of human societies \cite{Bainbridge2016,Burrows2012,Ferrari2008}.\par Since 2001, a number of other government initiatives \cite{frombrain,nesd,Prabhakar2017}, as well as other private initiatives \cite{Constine2017,kernel,neuralink}, have spawned, trying to address the technical and economic challenges of this subject in the agenda of their fundamental research programs \cite{Grau2014,Pais-Vieira2013}. Hereafter, some academic groups and non-governmental organization institutions, even dedicated ones \cite{miri,cser,fhi,ieet}, started to pronounce themselves on the ethics and consequences of converging technologies for human enhancement \cite{Attiah2014,Bainbridge2006,Bostrom2013,Clausen2009,Funk2016,Khushf2007}. As Ferrari \cite{Ferrari2008} already discussed, it’s interesting to note how differently the European Commission (EC) decided to frame this topic compared to the U.S. NSF’s original position. In 2004, as a response to the NSF’s definition of NBIC convergence, a \emph{High Level Expert Group} (HLEG) constituted by the EC, proposed in \emph{“Foresighting the New Technology Wave”} the notion of CTEKS (Converging Technologies for the European Knowledge Society) \cite{Nordmann2004}, which encompasses a much wider set of converging disciplines, including social sciences, policy, and philosophy. In this way, it emphasizes the parallel evolutionary path that, according to the EC’s HLEG, technology and society should take in order to account for the many ethical challenges inherent to such technological advancements.\par Although our position considerably stands with the wider framework proposed by the EC’s HLEG (rather than with other reductionist positions centered on a mere technological solutionism), the societal implications of such technologies embrace a too-broad area, and their exhaustive coverage is outside the scope of the present manuscript. Thus, here we want to begin by focusing on two specific aspects raised by such developments:
\begin{itemize}
  \item First, we reflect on how these converging technologies are reshaping our biological identity, as well as notions such as \emph{agency} and \emph{intentionality}. Therefore, how the traditional ontological and epistemological division between natural or artificial systems no longer applies, and what does it entails for the governance of human-systems.
  \item Secondly, we ask how these converging technologies are influencing the evolution of social structures, and support---or threaten---the implementation of more effective governance models in this unique moment of history. 
\end{itemize}
We will discuss these two independent---nonetheless connected---questions and conclude on the underlying implications of converging technologies both in our biology and in our society. From here on out, we will refer to the confluence between human and machine as the \emph{human-machine convergence}.

\section{Technological evolution: shaping a new biological identity}

\subsection{There is plenty of room at the bottom}
Physicist Richard Feynman \cite{Feynman1960} has speculated that there should be \emph{“plenty of room”} for an unavoidable development in which we will create smaller and smaller machine tools. Nowadays, manufacturing allows operation on a nanoscopic scale with considerable precision. These tools, according to Feynman, \emph{“might be permanently incorporated in the body to assist some inadequately-functioning organ”}. Presently, the tremendous work of scientists has kindled a positive attitude toward this speculation. \par The extraordinary abilities of synthetic and artificial machinery have enabled operations at the \emph{micro-} and \emph{nano-} scale, for instance at the level of single genes with remarkable precision \cite{Cong2013,Nelson2005}. Recent advancements in the production of reliable ultra-high frequency nano-antennas are providing another potential way to create implantable, controllable chips that could read and stimulate single neurons \cite{Nan2017}. Moreover, one of the highest barriers towards reliable brain implants has been recently overcome, providing us with a glial scar–free integration patch on a living brain \cite{Luan2017}. In particular this finding has inspired several entrepreneurs, among whom is the widely-known CEO of \emph{Tesla Motors, Inc.}, Elon Musk, to launch an endeavor to commercialize an ultra-high bandwidth brain-machine interface to connect humans and computers \cite{neuralink}.\par The technology of miniaturization gives rise to potential real-time data transfer in wearable, implantable and injectable devices, profoundly reshaping the way we behave, reason, and interact. As a consequence, the emergence of miniaturized devices may speed up the convergence between biological systems, such as human bodies, and their extensions (artifacts), thus creating a context where the self-artifact integration will fundamentally alter the biological notion of individuality and agency.

\subsection{Social reality and virtual reality}
In the last decade, technological progress has been considerably shaped by our needs for interpersonal interactions, mainly for social networking and information exchange \cite{Katz2002,Sarker2003}. Meanwhile, the drive for technological innovations has influenced our needs, extremely amplifying our innate reciprocal mirroring \cite{Danchin2004}, creating new behaviors on both individual \cite{Oulasvirta2012,Sparrow2011}, and collective \cite{Kramer2014} levels. The Internet is increasingly affecting how we, both as individuals and as a society, create, store, and retrieve information \cite{Garcia-Gavilanes2017}. In fact, the growing rate of information exchange is speeding up the augmentation and redefinition of our social bonds \cite{Ling2005}. The development of virtual reality has also changed the conventional view of interaction between humans and the environment \cite{Ohta2014,Taylor2002}. With advancements in the interdisciplinary field of \emph{Human Computer Interaction} (HCI), applications of virtual and augmented reality would not only imply that the users can break physical constraints, but also achieve a real-time interaction with other individuals \cite{Renard2010,Starner2017}.\par Blurring the boundaries between real and virtual environments could have a historically significant impact on identity on both personal \cite{Taylor2002} and interpersonal levels \cite{Holmes1997}.

\subsection{From artificial life to brain simulation}
Before looking into the reciprocal organization of humans and machines, one should understand how machines or software in computer simulations can physically self-organize and reproduce themselves as their biological counterpart. Von Neumann \cite{Neumann1951} was the first to formalize this question from a logico-mathematical standpoint, and detailing an elaborate solution regarding the design criteria and specifications \cite{VonNeumann1966}. Langton \cite{Langton1997} developed this schema and posed the concept of \emph{artificial life}, followed by theories of self-organization and emergence in nonlinear dynamical systems \cite{Waldrop1993}. These efforts are continuing nowadays as we move in understanding the fundamental laws that determine the organization of the universe from subatomic formation of matter to the emergence of living beings \cite{Schmickl2016}.\par Artificial life is an interdisciplinary subject that is not reducible to either theoretical biology or computer science; in fact, it draws fundamentally upon both and raises questions pertaining to life and its relation to information \cite{Boden1996,Steels1993}. Recent efforts to accurately create a computational replica of animal nervous systems \cite{Sporns2005} have opened the possibility for a deeper understanding of how the brain works in a complex environment. Some of these approaches approaches focus on the biological foundation of the brain \cite{Szigeti2014}, while others attempt to simulate the cognitive and behavioral processes observed with psycho-biological measurements to derive ideal machinery to replicate cognitive and behavioral functions \cite{Djurfeldt2008,Markram2006}. Which of these approaches---if any---will prove to be successful is still under debate; however, it is clear that there are great funding efforts towards machine learning, artificial intelligence, and massive neuro-computational simulations\cite{Guo2016,Schmidhuber2015}.

\subsection{Human and machine: an ontological and epistemological issue}
The increasing symbiotic relationship between humans and artificial artifacts has the potential to reshape the fundamental anatomy and physiology of our species, questioning the traditional ontological distinctions of human and machine.\par Moreover, what appears of special interest in the human-machine convergence is not so much the component itself (whose development is largely driven by information and communication technology), but rather the influences (i.e \emph{transformation processes}) of such components on our social and environmental interactions (i.e.\emph{relations}). Since evolutionary theories have always avoided any teleological approach \cite{Mayr1961}, we are now facing an etymological issue to say the least: \emph{machines} are technology, and as such, they have a purpose of use, while humans as biological and autonomous agents do not comply to such formalization.\par In fact, in the light of the human-machine convergence, the divide that characterizes the standard definition of both human and machines (in terms of structures, functions, and dynamics) no longer applies. The technologies moving us towards confluence with machines represent a critical point for scholars. Both natural and social sciences need to rethink what kind of subjectivity and metaphoricity are bound to their ontology and epistemology.
If machines are increasingly seen as an extension of our capabilities, while biology as a particular type of machine (e.g., in synthetic biology), we propose to explore novel semantics that still consider the existing distinctions between human and machines but that simultaneously take into account the functional analogies and the structural similarities between humans and machines.\par Moreover, we recognize that life tend to keep similar structures and dynamics throughout the emergence of more complex behaviors \cite{Bar-Yam1997}. In fact, at the social level, human civilization and its societies can even be considered as units or organisms capable of protecting their components and responding to changing environmental demands, what cyberneticist Heylighen calls a \emph{“super-organism”} \cite{Heylighen1996}. Furthermore, the expanding ability of internet-mediated organization is allowing for more complex and larger social groups to emerge. Additionally, the development of cheap, diffused, and connected sensors \cite{Cocchia2014,Gubbi2013,Zanella2014}, alongside the increasing miniaturization of the systems component discussed above, will---most likely---enable the emergence of a \emph{super-organism} made of humans and \emph{"things”} connected together by technological means.\par As we explore exterior, interior, and digital spaces \cite{Heylighen1996,Taylor2002}, complexity approaches provide a more comprehensive bodywork that allows semantic re-adaptation of the biological notion of the natural and artificial, and also of the individual and the collective \cite{Ball2012,Latour1993,Luisi2016}.\par In the context of Systems Design, from where our research draws as well, several scholars before us \cite{Backwell2009,Backwell2011,Fischer2006,Maturana1997} have already proposed non-reductive frameworks with the aim of embracing the emergent and unpredictable consequences of design solutions, acknowledging that \emph{“future uses and problems cannot completely [be] anticipated at [the] design time”} \cite{Fischer2006}. Such conceptualizations are known as \emph{"metadesign"}, a notion that was first introduced as an industrial design approach to Complexity Theory and Information Systems by Dutch designer Andries Van Onck \cite{VanOnck1965}. Metadesign offers different approaches to the design of systems that \emph{“include a coadaptive process between users and a system”} \cite{Fischer2006}. Furthermore, it attempts to solve the conundrum of certainty between roles, actions, and descriptors that revolve around the increasingly intimate relationships between humans and the designed systems.
\par In a recent article, on the implications for Design and Design education in the \emph{“Age of Hyperconnectivity”}, Iaconesi \cite{Iaconesi2017} concludes:\emph{“these [..] configurations of power schemes, practices and behaviors are at the border of what is assessed by laws, regulations, habits and customs. They are [..] new, unexpected, unforeseen, unsought. To confront with these issues, approaches which are trans-disciplinary are needed, because no single discipline alone is able to cover all of the knowledge, attitude, perspective which are needed to grasp and understand them. [..] In designing these ecosystems to confront with these issues it is necessary to make every possible effort to clearly and transparently define the boundaries of public, private and intimate spaces [..] as well as considerations that regard current business models, legislations, human rights, and (often national and international) security. There is no simple way to confront with this type of problem. [..]”}.
\par Certainly, we still lack a coherent epistemological and ontological framework that encompasses the specificities of both natural and engineered systems in their increasing intertwined relationship. Whether providing this framework is reasonably possible or not is a highly debated matter. Nonetheless, we encourage further exploratory studies along these lines.

\section{Social Operating Systems}
\epigraph{“[..] In time, those Unconscionable Maps no longer satisfied, and the Cartographers Guilds struck a Map of the Empire whose size was that of the Empire, and which coincided point for point with it. The following Generations, who were not so fond of the Study of Cartography as their Forebears had been, saw that that vast map was Useless [..] In the Deserts of the West, still today, there are Tattered Ruins of that Map, inhabited by Animals and Beggars; in all the Land there is no other Relic of the Disciplines of Geography.”}{--- \textup{Jorge Luis Borges}, Del rigor en la ciencia}

\subsection{Co-existence in a shifting and multipolar world}
The depletion of life-supporting resources, the decreased biocapacity of the planet, the increased risks of extinction for our species \cite{Barnosky2012,Rockstrom2009}, the concentration of wealth and the consequent global increase in inequality \cite{Davies2008}, the increase of global mental health disorders \cite{Collins2011}, the last two being intimately related \cite{Tanaka2017}), and the spreading of unstable regions that host autocratic governments, religious wars and some violent separatist movements \cite{Burrows2012}, are all issues that highlight the urgent need to improve our understanding of how cultural diversity can effectively co-exist, avoiding widespread social unrest and accelerated ecological transitions.\par Beyond the widely-adopted \emph{Human Development Index} (HDI), various efforts to provide more comprehensive measurements of developments have been already proposed. For instance, the \emph{inequality-adjusted HDI} \cite{Neumayer2001} additionally covers life expectancy, educational attainment, and per capita GDP. As an evolution of the inequality-adjusted HDI, some scholars began to question the role of cultural diversity in the context of an increasingly globalized world, expressing different scopes and identities \cite{Fukuda-Parr2004}. A variety of studies \cite{Alesina2005,Goren2013} agree that respect of diversity is key to a country’s development, yet sustaining peace represents a delicate process with non-negligible political and economic costs, as governments have to manage the demands for country resources of diverse competing groups.\par Furthermore, the advent of predictive frameworks for policy simulation leveraging large sets of data are widening, with some threats, the decision-making and operative capabilities of our species \cite{Bollier2010,Box2013,Edmonds2013,Iaconesi2017,Kim2014,Lafuerza2016}. Indeed, such potential raises important debates on how our institutions should leverage on an increasing amount of data to manage human activities on Earth avoiding to loose the grasp on the algorithmic grounding of our own socio-economic system \cite{Morozov2014,OReilly2013,Pasquale2015}.

\subsection{Sensing, foresight and governing the commons}
Statistical and applied probabilistic knowledge constitutes the fundamentals of risk-assessment and forecasting in our complex society \cite{Kyburg1991,Mantegna2005}. It is widely acknowledged that the near future for our species is raising compelling challenges \cite{Burrows2012,Stern2007,Donges2017}. Moreover, we already know that—to a certain extent—we have the ability to adapt our behavior and our physiology to an ever-changing environment \cite{Karatsoreos2011,Lupien2009}, and that conservation practices may also develop incremental elaboration of environment knowledge \cite{Berkes2008}. We, therefore, ask how our scientific forecasting approaches may detect early warning signs that could suggest safer trajectories of development for our activities on Earth. Conversely, from a socio-technical standpoint it still remains unclear how such approaches could and should effectively inform both traditional political institutions, as well as emerging power structures to reach renewed global institutional agreements.\par
Nonetheless, the aforementioned societal potential still raises compelling questions on which collective decisions are desirable in order to address the major ecological and social challenges of our times while guaranteeing a sustainable and socially fair integration of ICT in the human evolution. It is certainly challenging to guarantee that insights that arise from large data set analysis represent as much as possible the broad cultural diversity that humans have, in order to reasonably avoid cultural biases and social inequality. In a scenario where technology could simulate large data sets to predict the outcomes of policy at the economic, political, and financial levels, it is important to avoid biases and assumptions that might narrow the palette of research and that might legitimize unrepresentative actions of public and private groups that own such technical systems \cite{Boyd2012,Mantegna2005}. As other scholars agree, in order to allow for inclusive and participatory decisions to take place, systems and methods of research have to be made accessible, usable, and performable by all parties involved \cite{Darking2008,Iaconesi2017,Iaconesi2013}.\par 

\subsection{The potential growing at the edge of the ecosystem}
Our call for a more participatory and inclusive cultural diversity in advanced research methods is not only prescribed by ethical, ideological, or political reasons. Instead, the main reason for our conclusion came from the observation that, despite the challenges of integration processes, diversity is a key factor of resilience when environmental and ecological conditions change \cite{Bellwood2003,Elmqvist2003}. Intuitively, diversity was proved to be a key factor of resilience, especially in human communities during local emergencies \cite{Newman2005}. Other successful recurrent behaviors can be found in nature where diversity and delegation are key to optimize systems for dynamic environments. For instance we could observe a persistent logic to delegate the response to most of the individual needs in collaborative living systems \cite{Nowak2010}. This strategy partially compromises individual autonomy, but it allows an individual agent to contribute to a global need to survive and replicate while simultaneously allowing it to autonomously explore more complex tasks that lead to reduced functional redundancy. Ultimately, this approach lead the whole species to cover more biological niches reducing the number of points-of-failure \cite{Bonabeau2000,Dussutour2004}.\par In relation to what previously argued on foresight approaches to policy making, we could ask: how such sensing capabilities would be enriched if a context for the co-existence of diverse actors is provided and maintained? In this regard, for instance, we could clearly envision an interesting interplay of enabling players such as transborder systems integrators like the many UN System agencies and their stakeholders in  public and private sectors. By enabling, sustaining and empower a wide set of agile local players, such powerful organizations could potentially favour the co-existence of a multipolar set of niches without necessarily reduce the palette under "one size fits all" policy frameworks.  \par 

\subsection{New value creation structures} 
The \emph{“Internet of Things”} (IoT) paradigm \cite{Cocchia2014,Gubbi2013,Zanella2014}, and one of its concurrent evolution named \emph{“the Quantified Self”} \cite{Swan2009}, refer to the pervasive and diffused use of sensors and computing devices to gather and analyze large sets of data from environmental and biological agents, respectively \cite{Rabaey2002}.\par Some scholars argue that, in the information society, the amplified capability to collect and analyze data (which as a consequence leads to optimization of productivity through knowledge engineering), enabled by the diffusion of improving digital technology infrastructures, is resulting in the emergence of interdependent self-organized systems of (mostly immaterial) relational production \cite{Arvidsson2011,Arvidsson2012}. In the Fourth Industrial Revolution \cite{Cummings2017,Schwab2017}, these systems are consolidating novel forms of value-creation that are conceived more around ethics, purpose, and common goals than around the mere exploitation of labor as it was in the industrial society \cite{Arvidsson2011,Arvidsson2012,Hagel2017}. The fact that immaterial production revolves around abundant resources should in theory allow for an easier concurrence of needs among different groups; a previously limiting factor \cite{Alesina2005,Goren2013} in former economic paradigms that were mainly relying on the accumulation of scarce material goods.\par It still remains unclear whether it is possible to leverage technological means and governance systems to identify individual and collective behaviors that prevent our species from reaching unsafe tipping points in the planet ecosphere, and that simultaneously balance interests among different cultures having divergent scopes. Generally speaking, many remain skeptical about the possibility of a total control of the \emph{emergent} through technological means. In particular, criticisms around knowledge engineering argue that it is---not only---inherently limited our ability to effectively forecast the future, but that assertive or even prescriptive attempts might be extremely dangerous since they hinder resilience in regard to unpredictable scenarios \cite{Hendry2014,Taleb2009}. Concurrently, other scholars argue that the access to the massive quantities of information produced by and about people, things, and their interactions might usher in privacy incursion by invasive marketing and oppressive regimes in favor of an increase of financial and political control \cite{DeFilippi2014,Morozov2012,Weber2010}.

\subsection{The governance of ICT}
Governance is defined as the capability of social structures to manage processes and decisions that seek to define actions, grant power, and verify performance \cite{Bevir2012}. The human-machine convergence is addressing the role of ICT in our evolution as a species and pushes a compelling need to define proper policy frameworks for a safe and neutral confluence between humans and machines \cite{Cheng2011,Hahn2006}. Such policies should take into account the gap of governance that we are experiencing at the global level, where many countries still have stalled governments that are not able to keep the pace of global technological development \cite{Burrows2012}. Here non-state actors that act on a local level are experimenting with novel forms of interaction and decision that could eventually be globally amplified by ICT-enabled systems, leading to a more plural leadership and more representative governance of emergent behaviors.\par 
The low barrier to access to ICT make such technologies ubiquitous, thus enabling mass-scale coordination across geographic boundaries with near-instantaneous responses. In the last decade we have witnessed the use of this capability to raise global attention to the need for socio-political change \cite{McCaughey2013}. Indeed, ICT enables individuals to organize around common ideas in the virtual space and carry out real world action. Arguably, the more widespread new communications technologies are, the more they are rapidly becoming a double-edged sword for current governance, either supporting democratic transition or reinforcing autocratic regimes.\par 
According to a forecast report of the U.S. National Intelligence Council (NIC)–\emph{Global Trends 2030: Alternative Worlds}\cite{Burrows2012}, in the next decades individual empowerment will mainly accelerate for the diffusion of new communication and manufacturing technologies, while social networking will increasingly enable citizens to organize around specific scopes and empower their actions to challenge governments and institutions. On the other hand––the U.S. NIC’s report continues---such technologies are providing governments the unprecedented ability to monitor their citizens. Citizens are now demanding neutrality, participation, open access, transparency, and decentralization. The U.S. NIC’s report concludes that governments that fail to open up, avoiding the emergence of this process, are likely to face instability in the middle-long term.
The U.S. NIC’s report predict that during the next 15 to 20 years, as power becomes even more diffused than today, a growing number of diverse state and non-state actors, as well as subnational actors such as cities, will play important governance roles in an increasingly multipolar world. In a recent commentary in \emph{Law and Political Economy}, Frank Pasquale argue that quasi-monopolistic digital platforms are already marking the shift from territorial to what he calls functional sovereignty, creating a new digital political economy \emph{“undermining the territorial governance at the heart of democracy”} \cite{Pasquale2017}.

\section{Common structure and dynamics of social and biological systems}
Technology has always taken inspiration from nature. Here we follow this trend by focusing on how natural self-organizing systems might enlighten the issue of governance in scalable human organizations.
\subsection{The nervous system: an insightful analogy to our social system}
Science relies on metaphors; analogies are parsimonious ways to convey complex ideas and an efficient solution to reach common understanding. The nervous system has thus throughout history been compared to many systems, most of the time artificial. Descartes compared our brain to a mechanical machine, Freud to a steam machine, and the whole cognitive science project—coined at the Macy conferences—took the computer as its leading metaphor \cite{Dupuy2009}. Each time, the brain was hence compared to the most advanced human technology. With the rise of the Internet, it is not surprising that neuroscientists are now heading towards the metaphor of networks for understanding the brain \cite{Sporns2011}. In the same way computers were associated with information theory, networks are going hand-in-hand with graph theory. This mathematical framework has proven to successfully describe common patterns on separate scales, as seen in the multiple burgeoning -omics fields in system biology: proteomics, genomics, connectomics, metabolomics, etc.\par To what extent can the graph formalism bring insights across these scales? At first glance, the development of graph theory supports the existence of common universal dynamics and structure \cite{Boccaletti2006,Strogatz2001}. This is firstly interesting from a pure ontological perspective since it allows a parsimonious account of phenomena, avoiding the issue of reductionism that leads to an explosion of postulated mechanisms for different systems and contexts. Another interesting consequence is the facilitation of interdisciplinary work by narrowing the barrier between ontologies across disciplines.\par A potential path for a human-machine convergence might also rely on the bio-inspired design and deployment of our technology. The way our brain processes information through the cooperation between distributed brain areas provides pragmatic solutions for the design of resilient and innovative social networks. A key property of the nervous system is its subtle balance between integration and segregation \cite{Tononi1994}. This is partially reflected in the brain’s anatomical structure, the human connectome, which presents both specialized regions (with specific cytoarchitectonic features) and long-distance connections between them \cite{Sporns2005}. Similar topology appears in social networks: the so-called \emph{"small world"} networks \cite{Strogatz2001}. Their first feature is that they are neither hierarchical nor random, and they have a few hubs strongly connected to many other nodes which are loosely connected. This allows the integration of information by transit through the hubs; like this, the information can traverse the entire network with a low number of connections (as opposed to a fully connected network). In nature, segregation occurs with the emergence of communities or \emph{clusters} in which the elements are more interconnected between each other than with the rest of the network \cite{VandenHeuvel2011}.\par So, why take inspiration from the brain if the same pattern is already emerging spontaneously on the social level?
\subsection{Connectivity, inequality, and neutrality}
If the patterns observed on the social levels are similar to those on other scales, that does not mean all these patterns are efficient or valuable.\par For instance, the \emph{“rich-club”} phenomenon refers to the tendency of dominant elements of a system to form tightly interconnected communities. This formation of dominant communities in a given network is present in the healthy brain \cite{VandenHeuvel2011}, and a connectivity imbalance has been associated with schizophrenia \cite{VandenHeuvel2013} and epilepsy \cite{Spencer2002}. On the social scale, socio-physicists also uncovered such a \emph{"rich-club"} community at the core of our financial system \cite{Varela2001} and conclude at its potential role in its intrinsic instability, leading to the systemic risk of triggering financial crisis \cite{Battiston2012}. If solutions have already been proposed to counterbalance this accretion of control \cite{Vitali2011}, intrinsic conflicts of interest interfere with the regulation of the financial system \cite{Saunders1990}. In game theory, multi-agent models demonstrate that regulation by \emph{defectors} is less efficient than enforcement by \emph{collaborators} \cite{Pereda2016}. Nevertheless, redistributing the regulatory power in the hands of collaborators is not a trivial task.\par
With the advancement of the Internet, much hope arose concerning E-democracy, the enhancement of citizen’s access to political processes and policy choices. The project nonetheless faced the challenge of designing commons able to release the civic energy of its citizens \cite{Bria2015,Levine2002,Vercellone2015}. E-democracy development is also connected to complex internal factors, such as political norms and citizen pressures \cite{Lee2011}. Iceland recently demonstrated the operationalization of E-democracy through a grassroots participation in the constitution-making process \cite{Bani2012}. This proves that non-structured, non-hierarchical involvement of ordinary citizens, with a strong use of Web 2.0 tools, can promote public participation in ongoing governance.\par While self-organization creates adaptability and innovation, unenforced or unenabled cooperation can have down-side effects, such as the tragedy of the commons \cite{Hahn1968}. Regulation may thus be required at a certain level; the challenge is to enforce such regulation while avoiding the concentration of power. \par At the individual level, game theory simulations demonstrate that the \emph{tit-for-tat} strategy better performs over the long term than pure cooperation or pure defection \cite{Axelrod1981}. In a distributed schema such strategy can take advantage of ICT with the management of reputation at the global level instead of the local neighborhood: the actors of the system are informed of who is defecting and can adjust their future interactions accordingly \cite{Scott2005}. To date, the centralization of media (e.g., 90\% of the US media is now controlled by only six companies, down from 50 in 1983) \cite{Bagdikian1983,Baker2006,Lessig2004} has biased such modulation of reputation because of the centralization of power and asymmetry of communication. A distributed governance hence demands a new form of media—distributed and more symmetric.
\emph{"The media is the message"}, as McLuhan \cite{McLuhan1962} said; the rise of social media successfully proved his vision of our society becoming a \emph{"global village"}. From the Arab Spring to new forms of whistleblowing, ICT has also given the opportunity to all citizens to share and spread information without centralized control. This diffusion of power has been increasingly enabled by the development of peer-to-peer (P2P) and cryptographic technologies, guaranteeing transparency and anonymity simultaneously. While still flourishing technologies, they trigger collective experimentations in all fields, from currency and economy to the management of knowledge and common goods. Initiated decades ago by the open source movement, science is now challenging its own structure through open access publishing and, more generally, open science \cite{Bartling2014}. This culture of openness advocates for open knowledge, collectively organizing, discovering, and sharing data, unbundled tools, and information. Despite those values and ideas not being new in science, ICT intrinsically facilitates those processes while engaging a larger public \cite{Kera2014}. As the work of Kera \cite{Kera2012,Kera2014} and Delfanti \cite{Delfanti2012,Delfanti2013} shows, from bio- to nanotechnologies, DIYbio and hackerspaces movements have also created decentralized, participatory, and design-oriented practices, leading to alternative grassroots research and development approaches with experimental forms of ethical deliberation and regulation.

\section{Conclusions}
The research and development endeavors around converging technologies for human enhancement are raising important debates on reshaping the semantic boundaries between humans and machines. The symbiotic relationship between human biology and artificial artifacts tends to weaken our traditional epistemological and ontological division between what we define as human or machine. Indeed, the human-machine convergence is a call for rethinking what kind of metaphoricity and discursivity are bound in the disciplines related to human biology and technological development. Such a critical point may pave the way for novel approaches to understand the many relations between humans, machines, and the life-planet system they progressively occupy \cite{Fukuyama2002}.\par The human-machine convergence addresses the role of ICT-enabled systems in our biological and social evolution and requires defining adequate policies and approaches for a safe, sustainable, and neutral technological development of human society. Such approaches have to primarily account for the rapid depletion of Earth’s life-support systems, which unambiguously represents an unsustainable path for the development of our activities on this planet.
In coming years, the increasing number of individual and non-governmental players will complicate decision and policy making on a global level. Traditional organizations—i.e. state and institutional actors—will thus need to account for the effect of diffused technological empowerment, growing multipolarism, and macroeconomic shifts.\par Despite a lack of consensus, people, governments, and institutions, which need to co-exist in a progressively challenging scenario, will have to incentivize the emergence of innovative governance structures. As of today, it is unclear how traditional political structures will cope with the current development of ICTs \cite{AnderssonSchwarz2017,Cohen2016,Kilpi2016}, especially with the unpredictable advancements of converging technologies for human enhancement. However, these governance structures will have to adapt fast enough to harness change addressing the growing risk of widespread turmoil.\par
Our perspective seems to support that a stable, fair, and unbiased global sociopolitical system may depend on the rigorous consideration of establishing a novel type of distributed \emph{"systems leadership"} \cite{Hagel2017}. This new global leadership should aim at the deployment of a continuous learning process at the architectural level by experimenting and adopting modular, interoperable, and accountable standardized components and interfaces to enable distributed P2P practices \cite{Mattila2015,Seppala2016}. Such systems would reduce \emph{“the cost of adaptive coordination”} of traditional sociopolitical apparati, and it will allow for emergent coordination mechanisms \emph{“without the need to continually exercise authority”} \cite{Sanchez1996} in the form of rigid and fragile coupling typical of traditional governance structures. The unprecedented global leadership we envision would probably be ignited and constituted by the massive mobilization of loosely coupled networks of public and private mixed actors, ranging from governments to corporations to third sector organizations, leveraging on the capabilities of the thousands of networks and communities spread around the globe, as well as on the collective intelligence of billions of individuals pursuing common sustainability goals. From a merely technical standpoint, the ability to \emph{pull} and steer a large number of multipolar actors around common goals on a global scale is today difficult but an unquestionably possible objective. In a historical moment where knowledge is increasingly commodified and where production is increasingly automated, value is primarily generated through relational capital.\par The future of our species might depend on integrating the delicate interdependence between sustainability priorities and our fundamental human needs. We believe that such collective learning systems \cite{Kilpi2017} should aim at the empowerment of loosely coupled and locally distributed actors \cite{Navajas2018}, relentlessly experimenting towards novel kinds of global leadership on media, infrastructures and contexts designed to safely enhance the quality and quantity of interactions among peers.\par If our analysis will prove to have some heuristic value in the design of socio-technical systems, then a broad question remains open. In what practical ways techno-social systems can effectively learn and adapt with respect to our understanding of geopolitical complexity and evolutionary ecology?
 
\section{Acknowledgements}
We thank Célya Gruson-Daniel, Seonhee Kim, Salvatore Iaconesi, and Andrea Raimondi for their comments on earlier versions of this manuscript. We also want to express our gratitude to Jocelyn Ibarra for her great help in reviewing the latest version of this paper, and Simone Cicero for the endless inspiration in consolidating some of the most fundamental thoughts behind this research. Finally, we would also like to thank the Society for Autonomous NeuroDynamics (SAND) for constructive discussions and support.

\medskip  

\renewcommand*{\UrlFont}{\rmfamily}
\printbibliography
\end{multicols}
\end{document}